# Enabling High-Bandwidth Coherent Modulation Through Scalable Lithium Niobate Resonant Devices

Sadra Rahimi Kari[1,*], Paolo Pintus[2,3], John E. Bowers[2], Matt Robbins[4], Nathan Youngblood[1,*]

[1]Department of Electrical and Computer Engineering, University of Pittsburgh, Pittsburgh, PA 15261, USA
[2]Department of Electrical and Computer Engineering, University of California Santa Barbara, Santa Barbara, CA, USA
[3]Department of Physics, University of Cagliari, Cagliari, Italy
[4]Cetus Photonics, Inc.
*Email: sadra.rahimi@pitt.edu, nathan.youngblood@pitt.edu

## Abstract

We present a compact, resonant-based coherent modulator on a thin-film lithium niobate (TFLN) platform, addressing the growing demand for high-speed, energy-efficient modulators in modern telecommunications. The design incorporates Mach-Zehnder Interferometers (MZIs) with a Gires-Tournois etalon in each arm with a modulation region of only ~80 μm, eliminating the need for traveling-wave electrodes and enabling compatibility with wavelength-division multiplexing (WDM). Experimental results demonstrate a modulation bandwidth of 29 GHz, while ensuring low optical loss and high scalability. Our architecture supports in-phase and out-of-phase modulation, enabling differential control of amplitude and phase for advanced modulation formats such as quadrature amplitude modulation (QAM). Compared to previous designs, our approach enhances throughput, modulation density, and scalability, making it ideal for applications in coherent communications and optical computing. By combining the advantages of the TFLN platform with innovative resonator engineering, this work advances the development of compact, high-performance modulators for high-density on-chip communication networks.

## Introduction

The exponential growth in data traffic driven by modern telecommunication applications has created an urgent need for increased data transfer rates[1,2]. This ever-growing demand necessitates the design of electro-optic modulators with low power consumption, small footprint, and high throughput[3]. Integrated photonics offer a promising solution to these challenges, as they enable the miniaturization and energy-efficient operation of critical components required for high-speed communications[4,5].

Coherent communication techniques have proved instrumental in unlocking the full potential of existing technologies for building high-throughput networks[6–9] and are becoming important for emerging applications such as optical computing[10–13] and RF signal processing[14]. These coherent communication systems utilize in-phase and quadrature modulators (IQ modulators), which can independently control the amplitude and phase of optical signals, enabling advanced modulation formats[15] and achieving high spectral efficiency[16,17].

Among materials exhibiting second-order nonlinearity, thin-film lithium niobate (TFLN) stands out as an excellent candidate for integrated photonics due to its high Pockels coefficient, which facilitates high-speed modulation with low optical losses[18]. The lithium niobate platform has been widely explored for the development of high-performance coherent modulators[19–23]. However, majority of these approaches require TFLN modulators with lengths spanning several millimeters[18,24] to achieve high modulation efficiencies, which poses significant challenges for achieving high-density on-chip communication. Recent work[25] has introduced a compact TFLN photonic crystal (PhC) IQ modulator compatible with CMOS voltages, supporting modulation speeds of up to a few GHz. However, this approach relies on high-Q cavities, which inherently limit the data rate and fail to fully leverage the high bandwidth potential of the lithium niobate platform. Furthermore, the design faces scalability challenges, as it utilizes resonance peaks within the resonator's stopband, making it challenging to cascade these resonators to enable wavelength-division multiplexing (WDM).

Here we present a coherent, wavelength-selective modulator comprised of a Mach-Zehnder Interferometer (MZI) with a Bragg grating Gires-Tournois etalon in each arm. Here, we expand on the work of Pan et al.[26] in which the authors demonstrate a high-bandwidth LNOI electro-optic modulator based on a Fabry-Pérot (FP) architecture. Our resonance-based design allows for a compact modulation region 80 µm in length and avoiding the necessity of traveling-wave electrodes. The small footprint of our resonators allows for the cascading of multiple resonator pairs within an MZI, each tailored for a specific wavelength. The scalability of this design, combined with the high modulation bandwidth inherent to the TFLN platform, enables the development of high-throughput IQ modulators, with the aim of meeting the growing demand for increased data transfer rates and compact, coherent signaling on-chip.

## Results

**Performance Analysis of Single Resonators**

Our device concept, illustrated in **Figure 1**, integrates two Gires-Tournois etalons within a Mach-Zehnder Interferometer (MZI). Compact electrode pads are included to facilitate differential modulation within this architecture. **Figure 1a-b** outlines the operational principle of the device under differential modulation. By controlling the relative polarity of the electrical signals applied to contacts $V_{S1}$ and $V_{S2}$, the device can achieve either in-phase or out-of-phase modulation, controlling the amplitude and phase of the transmitted light, respectively.

For in-phase modulation (**Figure 1e**), the applied RF signals are in phase, generating external electric fields in opposite directions. One field aligns with the extraordinary (eo) axis of the lithium niobate (TFLN) crystal, causing a decrease in the refractive index (-Δn), while the other aligns opposite the eo axis, resulting in an increase in the refractive index (+Δn). This leads to a blue shift in one resonator and a red shift in the other, as depicted in **Figure 1a**. The separation of the resonances introduces a phase difference between the arms of the MZI, producing pure amplitude modulation. Assuming symmetrical differential modulation, where both resonators experience equal modulation, the output phase remains constant during in-phase operation.

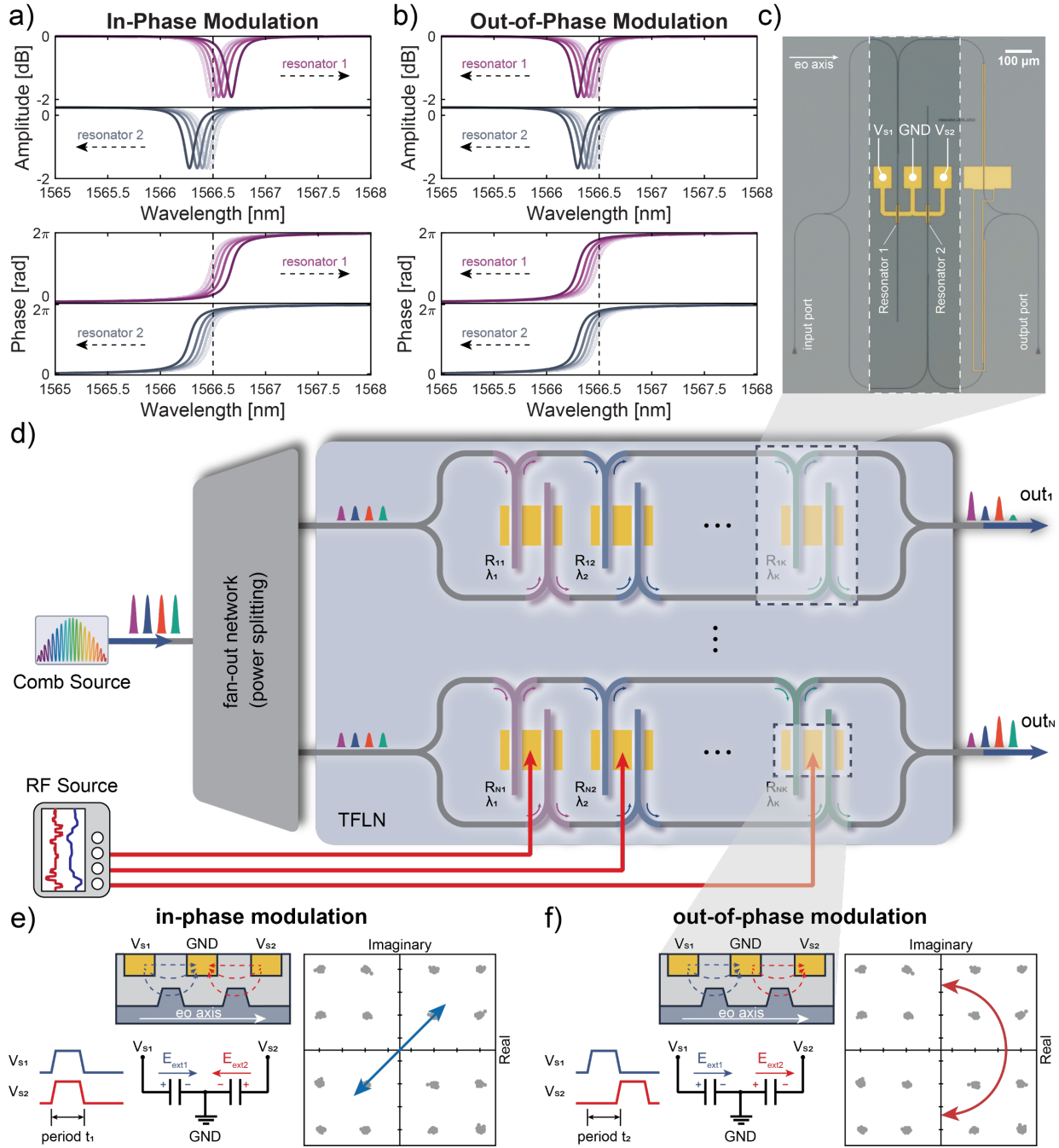

**Figure 1: Overview of proposed resonant modulators for coherent modulation. a-b)** Amplitude and phase responses of resonators during in-phase **(a)** and out-of-phase **(b)** modulation which results in changes in the amplitude **(a)** and phase **(b)** of the output signal, respectively. **c)** Optical microscope image of an MZI integrated with a single resonator pair. Modulation is achieved by providing differential signals to the contacts $V_{s1}$ and $V_{s2}$ **d)** Schematic of future device to increase throughput by leveraging WDM and SDM. **e, f)** Working mechanism of coherent modulator during differential modulation. The direction of generated external e-fields relative to eo-axis of TFLN crystal can be modified to perform amplitude **(e)**, and phase **(f)** modulations.

Conversely, for out-of-phase modulation (**Figure 1f**), the electric fields point in the same direction, causing the resonant wavelengths of both resonators to shift in the same direction, as shown in

**Figure 1b**. Under this mode, the relative phase difference between the arms remains constant, resulting in no change to the output amplitude. However, the phase of the output signal varies depending on the degree of modulation.

A microscope image of one such fabricated modulator with a single resonator pair is shown in **Figure 1c**. Samples were fabricated at Luxtelligence[26] using their open-source process design kit (PDK)[27]. The compact design of the resonator pair, featuring a modulation length of just 80 μm, ensures excellent scalability. Our vision for scaling this modulator platform is illustrated in **Figure 1d**, where wavelength-division multiplexing (WDM) is achieved by cascading multiple resonator pairs, each tuned to a specific frequency. Additionally, throughput can be further enhanced by employing spatial multiplexing, in which multiple copies of the modulators operate simultaneously. For a system with $K$ channels and $N$ modulators, each capable of encoding $2^B$ bits per symbol, the output bit rate can be expressed as $2^B \times N \times K \times f_{3dB}$. Leveraging the high modulation bandwidth of TFLN devices, our proposed system has the potential to achieve data rates in the Tbits/s range.

To minimize insertion loss while maintaining high transmission for off-resonant wavelengths, we leverage mode multiplexing where integrated Gires-Tournois etalons are implemented using phase-shifted Bragg gratings[28]. These gratings consist of a longer fully reflective section and a shorter partially reflective section, as shown in **Figure 2a**. Because we are using a phase-shifted Bragg grating design, the input optical $TE_0$ mode is converted to $TE_1$ upon reflection. This reflected $TE_1$ mode is then converted back to the fundamental $TE_0$ mode after passing through the output port of an adiabatic multi-mode coupler.

To better understand the impact of key design parameters on resonator performance, we fabricated multiple single resonators with varying corrugation widths ($W$) and lengths of the partially reflective mirrors ($L$). This systematic variation allows for detailed performance analysis across different configurations. Key metrics for optimization include optical bandwidth of reflection and quality factor ($Q$) of the resonator. An ideal device should have a sufficiently large optical bandwidth to accommodate the desired number of frequency channels ($K$) within its passband. The $Q$ factor is equally critical and must be carefully selected, as it directly influences the modulation bandwidth of cavity-based devices. The cutoff frequency ($f_{3dB}$) of our resonators is determined by a combination of the $RC$ constant of the electrodes and the photon lifetime ($\tau_p$) of the optical cavity[29].

$$\frac{1}{f_{3dB}^2} = \frac{1}{f_{\tau p}^2} + \frac{1}{f_{RC}^2}, \qquad (1)$$

where

$$f_{\tau p} = \frac{1}{2\pi \tau_p}, \text{ and } \tau_p = \frac{Q_t}{W_0} \qquad (2)$$

Equivalently,

$$\frac{1}{f_{3dB}^2} = \left(\frac{2\pi Q_t}{W_0}\right)^2 + \frac{1}{f_{RC}^2} \qquad (3)$$

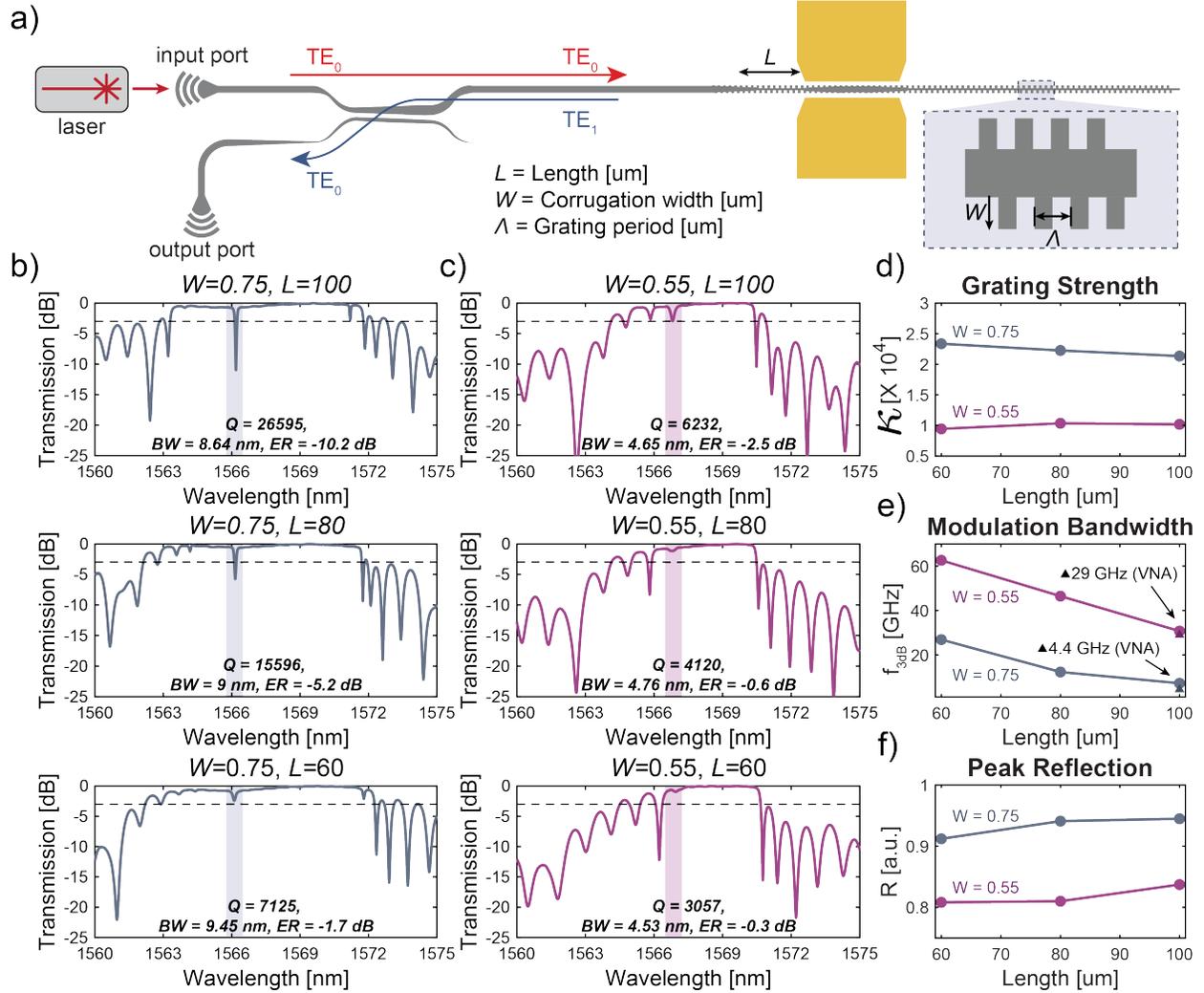

**Figure 2: Performance of integrated Gires-Tournois resonators. a)** Schematic of the device designed using a single Gires-Tournois etalon and adiabatic couplers to perform mode conversion. **b-c)** Frequency spectrum of fabricated devices with different lengths for corrugation widths of 0.75 μm **(b)**, and 0.55 μm **(c)**, respectively. **d)** Coupling coefficient of designed resonators extracted from experimental results shown in **(b)** and **(c)**. **e)** 3 dB modulation bandwidth of our resonators calculated using the $Q$-factor. **f)** Peak reflection values used in computational modeling of resonators.

Here, $Q_t$ represents the total quality factor of the resonator, and $W_0$ denotes the optical frequency. In our design, the $RC$ time constant of the electrodes has minimal influence on $f_{3dB}$, as the electrodes are very short and primarily capacitive. Consequently, the modulation bandwidth is predominantly determined by the $Q$ factor of the cavity, with which it has an inverse relationship.

The experimentally measured optical spectra of these resonators are shown in **Figures 2b** and **2c**. The results indicate that decreasing the resonator length ($L$) leads to a reduction in the $Q$ factor and extinction ratio ($ER$), while increasing the optical bandwidth ($BW$). Conversely, reducing the corrugation width ($W$) results in a decrease in $Q$ factor, $ER$, and $BW$. Careful selection of $W$ and $L$ is therefore critical to achieving sufficient optical bandwidth and a low enough $Q$ factor to accommodate the desired modulation bandwidths.

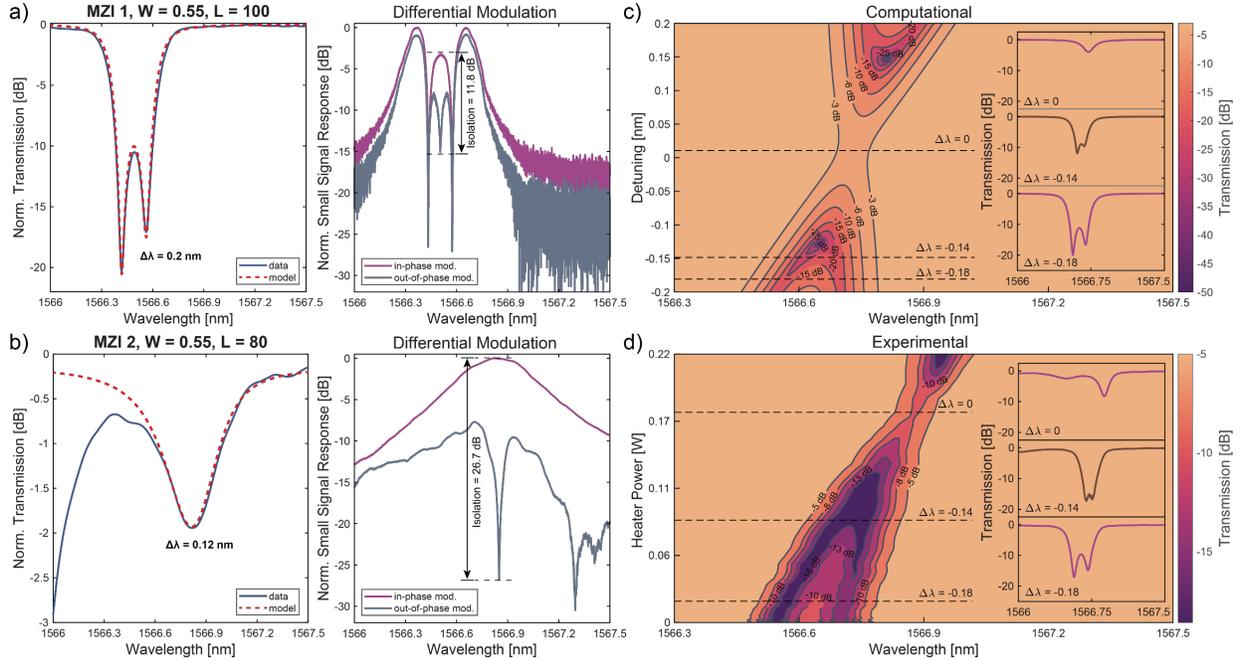

**Figure 3: Effects of resonance detuning in small signal response of coherent modulators integrated with a single resonator pair. a)** Frequency spectrum and small signal response of a MZI with corrugation width of 0.55 μm and length of 100 μm. Results show an isolation of 11.8 dB between in-phase and out-of-phase modulation for a resonator pair with 0.2 nm resonance detuning. **b)** Frequency spectrum and small signal response of a MZI with corrugation width of 0.55 μm and length of 80 μm. Results show a large isolation of 26.7 dB between in-phase and out-of-phase modulation for a resonator pair with 0.12 nm resonance detuning. **c)** A computational 2D heatmap of the $MZI_1$ transmission is presented as a function of detuning, with the transmission spectra highlighted for specific detuning values of $\Delta\lambda = 0, -0.14, -0.18\ nm$. **d)** An experimental 2D heatmap of the $MZI_1$ transmission is shown as a function of heater power, corresponding to detuning, with transmission spectra highlighted for specific detuning values of $\Delta\lambda = 0, -0.14, -0.18\ nm$.

Our fabricated devices exhibit an insertion loss of approximately 1 dB, while the measured back-reflection at the resonance wavelength is around -5 dB as measured at the input port (see Supplementary). Furthermore, using computational models and the extracted Q factor of the fabricated devices, we estimated the modulation bandwidth of the resonators, with the results presented in **Figure 2e**. Experimental validation was performed using a vector network analyzer (VNA), which demonstrated that the measured results closely align with the estimated values within an acceptable margin of accuracy. The VNA measurements were performed on devices with configurations of (W=0.75, L=100) and (W=0.55, L=100).

To gain deeper insight into the performance of these devices, a comprehensive computational model was developed based on the experimental results. This model allowed us to extract key parameters such as peak reflection (**Figure 2f**), grating strength (**Figure 2d**), and loss. For additional details on the modeling process, please refer to the supplementary information.

### Design and Performance of Coherent Modulators

Etalons with the same design parameters as the single resonators from **Figure 2** were embedded into two arms of an MZI, as shown in **Figure 1c**. Differential modulation for these MZI structures was achieved by applying RF signals to contacts $V_{S1}$ and $V_{S2}$ (500 kHz with an amplitude of 10 V for lock-in measurements). For in-phase modulation, the phase delay between the two signals was set to zero, while for out-of-phase modulation, a 180-degree phase delay was introduced. The output signals were captured using a lock-in amplifier to demodulate the signal and record the small-signal response of the modulators. **Figures 3a** and **3b** illustrate the normalized transmission (in the absence of modulation) and the normalized small-signal response during in-phase and out-of-phase modulation for two MZI devices: $MZI_1$, integrated with a resonator pair characterized by $W = 0.55$ μm and $L = 100$ μm, and $MZI_2$, integrated with a resonator pair characterized by $W = 0.55$ μm and $L = 80$ μm, respectively. Based on the normalized transmission data from the two MZIs, a computational model was developed for each to gain a deeper understanding of the modulators' performance. By fitting the measured spectra to our analytical model, we calculated resonance detunings of 0.2 nm and 0.12 nm for $MZI_1$ and $MZI_2$, respectively. Since the resonator pairs within each MZI were designed to be identical, the observed resonance detuning arises from fabrication variations. Furthermore, comparing the small-signal responses of the two MZIs reveals significant differences in their frequency spectrum profiles, primarily due to the different detuning values. Despite the presence of resonance detuning, both devices demonstrated significant isolation (11.8 dB for $MZI_1$ and 26.7 dB for $MZI_2$) between in-phase and out-of-phase modulation, indicating the ability to control phase and amplitude independently at resonance. The small-signal response captures only changes in the amplitude during differential modulation, so having a high isolation ensures that output amplitude varies during in-phase modulation, while remaining nearly constant during out-of-phase modulation. These experimental results strongly validate the operational principles of these modulators introduced earlier.

Resonance detuning not only affects the overall profile of the small-signal response but also impacts the isolation and probing wavelength of this response. These devices are highly influenced by fabrication variations, resulting in sensitivity to detuning. While some of this variability can be reduced by placing the resonator pairs in close proximity to one another[30], alignment of the two resonances is crucial to ensure reliable and predictable performance during modulation. To address this, we deposited (Ti/Au) heaters on top of the fully reflective mirror which allows for precise control of resonance detuning. **Figure 3c** shows the expected change in the transmission spectra of $MZI_1$ as a function of detuning based on our analytical model. At zero detuning ($\Delta\lambda = 0$), the transmission spectrum exhibits a small dip at the resonance wavelength, indicating minimal insertion loss of the modulator. Experimental results (**Figure 3d**) which use a thermo-optic heater to align the resonances show good agreement with the analytical model in **Figure 3c**. For more detailed information regarding the effects of resonance detuning on the small-signal response, please refer to the supplementary information.

One of the key advantages of our approach is the ability to achieve large isolation and maintain a stable amplitude response even with small phase shifts applied to each resonator. The relatively small phase shift during in-phase modulation results in sufficient separation between resonators, leading to a significant relative phase difference between the MZI arms. Computational modeling shows that pulses with an amplitude of 10 V produce a phase shift of approximately $\frac{\pi}{90}$ per resonator (see supplementary information). As a result, these devices remain fully operational at lower modulation efficiencies, eliminating the need for long modulation regions and bulky traveling-wave electrodes that limit device scalability. This has enabled the design of shorter electrode lengths (80 μm) for modulation, enhancing the compactness and scalability of our devices.

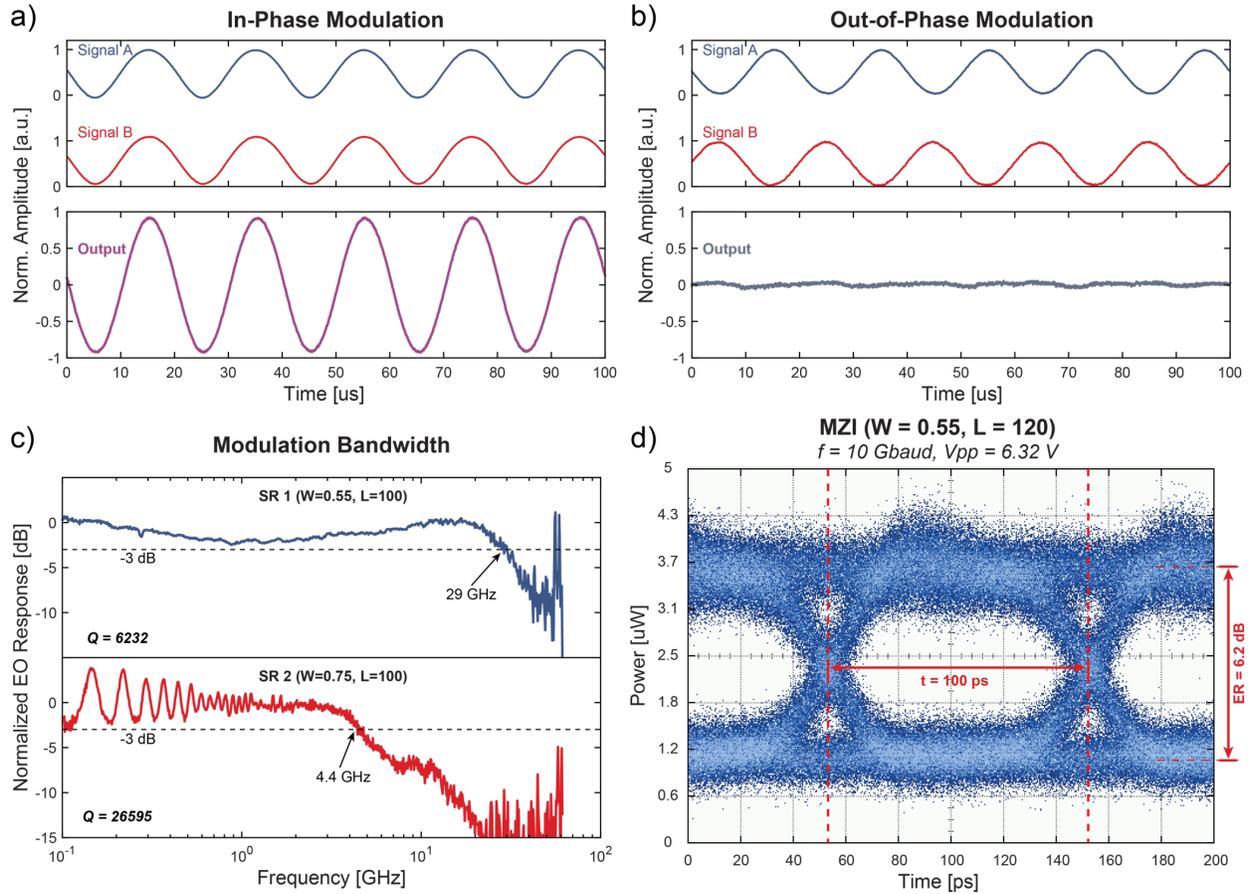

**Figure 4: High-speed performance of resonance-based modulator during coherent modulation. a-b)** Electro-optic modulated output signal of $MZI_1$ modulator during in-phase **(a)** and out-of-phase **(b)** modulation. **c)** Modulation bandwidth of single resonators with different corrugation widths of 0.55 um (blue), and 0.75 um (red), which resulted in *Q*-factors and 3dB bandwidths of $Q = 6{,}232$, $f_{3dB} = 29$ GHz, and $Q = 26{,}595$, $f_{3dB} = 4.4$ GHz, respectively. **d)** Eye diagram of the modulator measured at 10 Gbaud shows low jitter and an extinction ratio of 6.2 dB.

**Figures 4a** and **4b** display the AC-coupled output of the modulator during in-phase and out-of-phase modulation, respectively. As expected, in-phase modulation results in significant amplitude changes, which correspond to amplitude modulation (**Figure 4a**). In contrast, during out-of-phase modulation, the output exhibits minimal amplitude variations, indicating phase-only modulation (**Figure 4b**). **Figure 4c** illustrates the modulation bandwidth of two single resonators with varying Q factors. As discussed earlier, devices with lower Q factors exhibit higher modulation bandwidths. **Figure 4d** presents a captured eye diagram of an MZI during differential modulation operating at 10 GHz, showcasing the high-speed performance of the device.

## Discussion

Our analysis of experimental data and extensive computational modeling reveals that the performance of the modulators is highly sensitive to resonance detuning, affecting both achievable isolation and the probing wavelength. Our integrated thin-film lithium niobate (TFLN) devices exhibit greater fabrication variations compared to SOI platforms, primarily due to the relative immaturity of the TFLN fabrication processes and the growth of TFLN wafers which was observed to vary in thickness by +/- 20nm across a 6-inch wafer. Consequently, our design is particularly susceptible to mismatches between resonance wavelengths. To ensure accurate operation and reliable predictions for these modulators, implementing a control mechanism to compensate for detuning is essential. In this work, we addressed this challenge by employing integrated heaters to correct for resonance detuning. However, low-loss phase change materials, such as $Sb_2Se_3$, could be a promising nonvolatile method for aligning resonances after fabrication[31].

The parameters of the resonator were initially selected based on a theoretical design with simulations involving sweeps of $W$ and $L$. However, insights from our experimental analysis allow us to refine expectations regarding these parameters. At a high level, for applications such as wavelength-division multiplexing (WDM) or multi-channel systems, the optical bandwidth of a single resonator must be sufficient to accommodate 4–8 channels. Bandwidth is fundamentally determined by $W$ and $L$, with a tradeoff between the $Q$-factor and modulation bandwidth: a lower $Q$-factor leads to higher modulation bandwidth but less channels. The desired performance can be achieved through optimization of the corrugation width ($W$) and length of the first resonator ($L$). We have demonstrated more than 20× improvement in bandwidth, achieving 29 GHz compared to the previous state-of-the-art value of 1 GHz using a similar approach[25].

To further enhance system throughput, in addition to optimizing the corrugation width ($W$) and resonator length ($L$), other strategies can be employed. Increasing the number of channels within a single modulator enables wavelength-division multiplexing (WDM), while scaling the number of modulators allows for spatial-division multiplexing (SDM). These approaches collectively contribute to a substantial increase in throughput. Our coherent design also leverages the unique advantages of the TFLN platform to enable seamless integration of IQ modulators for QAM generation. Quadrature amplitude modulation (QAM) is a widely used scheme in high-speed optical communications, combining amplitude and phase modulation to transmit more information per symbol[32,33]. In-phase (I) and quadrature (Q) modulators play a critical role in QAM systems by enabling independent control of the amplitude and phase of the optical signal[24,34], which facilitates the generation of complex modulation formats[35–37]. TFLN-based IQ modulators have shown superior performance compared to traditional designs, offering support for high data rates while simultaneously meeting the requirements for low loss, low drive voltage, and large bandwidth[24]. Furthermore, advancements in modulation schemes, such as asymmetric modulation, have the potential to facilitate the generation of standard QAM constellations at higher rates than previously demonstrated in [25] using resonant-based coherent modulators. Preliminary analysis suggests that implementing 16-QAM is feasible using our device, please refer to supplementary information for additional details on QAM implementation.

Finally, these coherent, WDM modulators are not only valuable for telecommunications applications but can also be utilized as inputs for coherent analog computing. Prior research[38] has demonstrated high-throughput dot-product operations on the TFLN platform using cascaded MZIs. However, their approach is limited in scalability and requires a complex process for performing multiplications between two negative inputs. By employing a crossbar array with dot-product unit cells[39] or other time-multiplexed architecture[40] on the TFLN platform and substituting conventional MZMs with the modulators presented in this work, both throughput and scalability can be significantly enhanced for computing applications. Having the ability to encode information not only in wavelength, but also in the complex plane gives additional benefits to optical accelerators[41,42]. Assuming a modulation bandwidth of 29 GHz observed in the device with a configuration of (W=0.55, L=100), symbol rates of up to $2 \times f_{3dB}$ can be achieved[43]. Consequently, our proposed design is capable of reaching symbol rates close to 58 Gbaud per channel. By utilizing a configuration with four wavelength channels (WDM) and four modulators (SDM), the system can achieve a total data rate of 3.7 Tbit/s for 16-QAM encoding. Based on the carefully estimated $f_{3dB}$ values in **Figure 2e**, the device with a configuration of (W=0.55, L=60) is capable of achieving a modulation bandwidth of 60 GHz, corresponding to a symbol rate of 120 Gbaud. Table 1 presents an estimate of the achievable throughput for our proposed system across different network configurations when operating at symbol rates of 58 and 120 Gbaud.

Table 1. Performance Estimation of Proposed Architecture

| Number of Channels (WDM) | Number of Modulators (SDM) | Bit rate for 16-QAM (Tbits/s) | |
|---|---|---|---|
| | | Operating at 58 Gbaud | Operating at 120 Gbaud |
| 1 | 1 | 0.232 | 0.48 |
| 2 | 2 | 0.928 | 1.92 |
| 4 | 4 | 3.712 | 7.68 |
| 8 | 4 | 7.424 | 15.36 |
| 8 | 8 | 14.848 | 30.72 |

## Conclusion

We have demonstrated a resonant-based modulator integrating Bragg grating-based Gires-Tournois etalon pairs for coherent modulation. Our design leverages differential modulation to achieve amplitude modulation with in-phase RF signals and phase modulation with out-of-phase RF signals, thereby improving both throughput and modulation density. Compared to prior work[25], our approach improves modulation speed by over 20×, while operating with lower modulation efficiencies and greater cascadability. Our devices have also achieved excellent performance in key metrics such as optical bandwidth and modulation bandwidth, which are critical factors influencing scalability and throughput. The current design can be readily scaled to achieve Tbit/s data rates by integrating wavelength-division multiplexing (WDM) and spatial-division multiplexing (SDM) within practical configurations. These innovations hold promise for advancing resonant-based modulators that are fast, efficient, and scalable.

## Methods

*Measurement Setup:*

A tunable fiber laser (Santec TSL-550) was utilized to generate input light with wavelengths ranging from 1500 to 1630 nm for the experiments. A sixteen-channel fiber array was employed to couple light into and out of the chip via on-chip grating couplers. A two-channel custom RF Probe (S-G-S) from GGB Industries with a bandwidth of up to 40 GHz was utilized for Data modulation. To facilitate the capture and real-time recording of optical data from the chip, the output of a fiber-coupled photodetector (Newport 2011-FC) was connected to a BNC-2110 data acquisition board from National Instruments. To capture the spectral response of the resonators, the wavelength of the tunable laser source was swept from 1500 to 1630 nm at a speed of 100 nm/s, and 100,000 samples were captured from the output photodetector by the DAQ board for the entire sweeping range.

For capturing the electro-optic response and bandwidth of LNOI modulators, which requires operation at higher speeds, a high-speed photodetector (Newport 1544-A) with a bandwidth of up to 12 GHz and a vector network analyzer (Siglent SVA1075X) with a bandwidth between 100 KHz and 7.5 GHz were utilized.

*Device Calibration:*

Our coherent resonant modulators consist of Mach–Zehnder interferometers (MZIs) incorporating Fabry–Pérot cavities in each arm. Prior to conducting modulation experiments, we performed a calibration process on individual resonator devices to assess their performance based on different design parameters. Each device underwent a wavelength sweep using a tunable laser source to capture the spectral response of the resonators. Measurements were taken simultaneously from one device port and one test port using two photodetectors. The analog outputs were then digitized and recorded with a National Instruments (NI) Data Acquisition (DAQ) board. Performance metrics, including Insertion Loss (IL), Quality Factor (Q), Extinction Ratio (ER), and Bandwidth (BW), were calculated for each device with different design parameters. The insertion loss of our resonators was observed to be approximately 1 dB.

For the modulation experiments, we selected an MZI design featuring resonators with relatively lower Q factors to enhance modulation speed. Each resonator within the MZI was modulated individually, and the output was captured using a photodetector and a lock-in amplifier. The lock-in amplifier was employed to demodulate the electrical signals applied to the resonators and to obtain precise output measurements.

Finally, we developed a computational model for the MZIs to validate the experimental results and to gain further insights into the performance of these modulators.

## Data availability

All data supporting this study are available in the paper and Supplementary Information. Additional data related to this paper are available from the corresponding authors upon request.

## Disclosures

M.R. and N.Y. have filed a patent application related to this work. The other authors declare no competing interests.

## Acknowledgements

This work was supported by the National Science Foundation under awards #2337674 and 2105972 and the AFOSR Young Investigator Award FA9550-24-1-0064. P. Pintus acknowledges the support of the Autonomous Region of Sardinia via the 'Mobilità Giovani Ricercatori (MGR)' programme of the University of Cagliari.## References

1. Winzer, P. J. & Neilson, D. T. From Scaling Disparities to Integrated Parallelism: A Decathlon for a Decade. *Journal of Lightwave Technology* **35**, 1099–1115 (2017).
2. Heidari, E. *et al.* Integrated ultra-high-performance graphene optical modulator. *Nanophotonics* **11**, 4011–4016 (2022).
3. Lee, B. S. *et al.* Scalable graphene platform for Tbits/s data transmission. (2020).
4. Koos, C. *et al.* Nanophotonic modulators and photodetectors using silicon photonic and plasmonic device concepts. in (eds. Witzigmann, B., Osiński, M. & Arakawa, Y.) 1009807 (2017). doi:10.1117/12.2256536.
5. Qi, Y. *et al.* Design of Ultracompact High-Speed-Integrated Lithium–Niobate Periodic Dielectric Waveguide Modulator. *Adv Photonics Res* **3**, (2022).
6. Pfeifle, J. *et al.* Coherent terabit communications with microresonator Kerr frequency combs. *Nat Photonics* **8**, 375–380 (2014).
7. Xie, C. & Cheng, J. Coherent Optics for Data Center Networks. in *2020 IEEE Photonics Society Summer Topicals Meeting Series (SUM)* 1–2 (IEEE, 2020). doi:10.1109/SUM48678.2020.9161052.
8. Li, S.-A. *et al.* Enabling Technology in High-Baud-Rate Coherent Optical Communication Systems. *IEEE Access* **8**, 111318–111329 (2020).
9. Harter, T. *et al.* Generalized Kramers–Kronig receiver for coherent terahertz communications. *Nat Photonics* **14**, 601–606 (2020).
10. Hamerly, R., Bernstein, L., Sludds, A., Soljačić, M. & Englund, D. Large-Scale Optical Neural Networks Based on Photoelectric Multiplication. *Phys Rev X* **9**, 021032 (2019).
11. Pai, S. *et al.* Experimentally realized in situ backpropagation for deep learning in photonic neural networks. *Science (1979)* **380**, 398–404 (2023).
12. Chen, Z. *et al.* Deep learning with coherent VCSEL neural networks. *Nat Photonics* **17**, 723–730 (2023).
13. Bandyopadhyay, S. *et al.* Single-chip photonic deep neural network with forward-only training. *Nat Photonics* **18**, 1335–1343 (2024).
14. Li, P., Dai, Z., Yan, L. & Yao, J. Microwave photonic link to transmit four microwave vector signals on a single optical carrier based on coherent detection and digital signal processing. *Opt Express* **30**, 6690 (2022).
15. Li, S.-A. *et al.* Enabling Technology in High-Baud-Rate Coherent Optical Communication Systems. *IEEE Access* **8**, 111318–111329 (2020).
16. Hirokawa, T. *et al.* Analog Coherent Detection for Energy Efficient Intra-Data Center Links at 200 Gbps Per Wavelength. *Journal of Lightwave Technology* **39**, 520–531 (2021).
17. Seiler, P. M. *et al.* 56 GBaud O-Band Transmission using a Photonic BiCMOS Coherent Receiver. in *2020 European Conference on Optical Communications (ECOC)* 1–4 (IEEE, 2020). doi:10.1109/ECOC48923.2020.9333218.
18. Wang, C. *et al.* Integrated lithium niobate electro-optic modulators operating at CMOS-compatible voltages. *Nature* **562**, 101–104 (2018).
19. Zhu, D. *et al.* Integrated photonics on thin-film lithium niobate. *Adv Opt Photonics* **13**, 242 (2021).

# Supplementary Information for Enabling High-Bandwidth Coherent Modulation Through Scalable Lithium Niobate Resonant Devices

Sadra Rahimi Kari[1,*], Paolo Pintus[2,3], John E. Bowers[2], Matt Robbins[4], Nathan Youngblood[1,*]

[1]Department of Electrical and Computer Engineering, University of Pittsburgh, Pittsburgh, PA 15261, USA
[2]Department of Electrical and Computer Engineering, University of California Santa Barbara, Santa Barbara, CA, USA
[3]Department of Physics, University of Cagliari, Cagliari, Italy
[4]Cetus Photonics, Inc.
*Email: sadra.rahimi@pitt.edu, nathan.youngblood@pitt.edu

## S1. Working Principle

To gain deeper insights into the operational mechanism of our Mach-Zehnder Interferometer (MZI) design for coherent modulation, **Figure S1a** illustrates three distinct operating regimes.

In the first scenario, under ideal conditions where the resonance wavelengths of the two resonators embedded in the MZI are perfectly matched, no modulation is applied. In this regime, the output phases of the resonators, denoted as $\varphi_1$ and $\varphi_2$, exhibit no separation (**Figure S1c**).

In the second scenario, assuming ideal conditions with symmetrical modulation, out-of-phase modulation results in identical shifts in the resonance wavelengths of both resonators. Consequently, $\varphi_1$ and $\varphi_2$ experience no relative separation; however, both phases shift compared to the no-modulation case.

In the third scenario, during in-phase modulation, the resonance wavelengths of the resonators shift in opposite directions. This divergence induces a separation between the output phases $\varphi_1$ and $\varphi_2$. The relative phase difference between the MZI arms, denoted as $\Delta\varphi$, is influenced by the operating regime. As depicted in **Figure S1b**, in-phase modulation uniquely affects $\Delta\varphi$, leading to changes in the amplitude of the output signal while leaving the output phase unaffected. Conversely, during out-of-phase modulation, $\Delta\varphi$ remains constant, resulting in an unchanged output amplitude while the phase of the output signal varies due to modulation.

As with any MZI designed for coherent modulation, our devices are biased at the quadrature point (**Figure S1d**) to ensure symmetrical and linear responses during differential modulation.

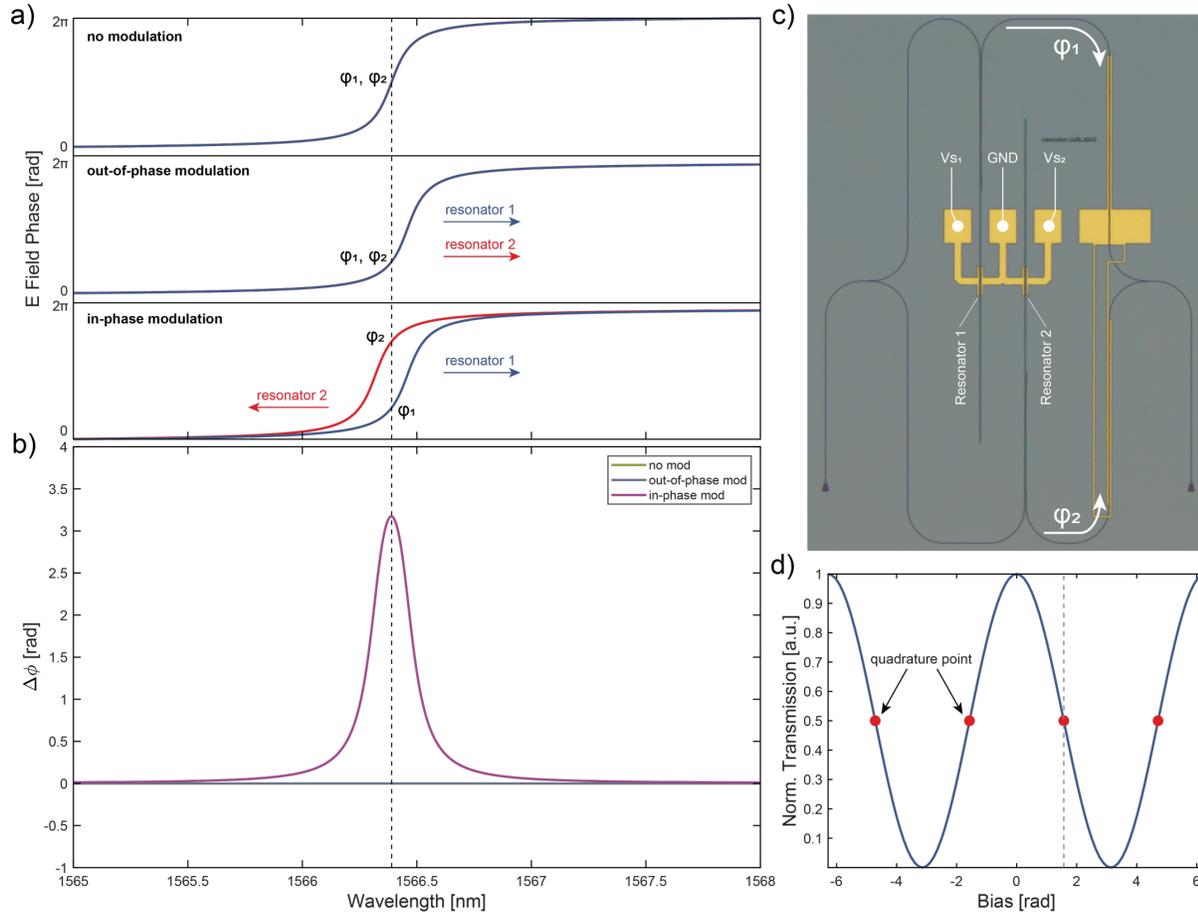

**Figure S1. Working principle of coherent modulators during differential modulation. a)** Output phase of resonators during no modulation, out-of-phase, and in-phase modulations. **b)** Relative phase difference between MZI arms during differential modulation. **c)** Optical microscope image of the fabricated coherent modulator. **d)** Transmission of the MZI as a function of its bias point.

## S2. Computational Modeling

Our resonators are designed using phase-shifted Bragg gratings, comprising a longer fully reflective section ($r_2 = 1$) and a shorter partially reflective section ($r_1$). **Figure S2a** illustrates a schematic of the single resonator design utilized for computational modeling, in contrast to the real device schematic presented in **Figure 2a** of the main text.

In the modeling process, the longer and shorter Bragg gratings are simplified and replaced with reflective mirrors. The longer grating, representing the fully reflective section, is assigned a reflection coefficient of $r_2 = 1$, while the shorter grating is characterized by a variable reflection coefficient, $r_1$. The cavity loss is denoted by $\sqrt{A}$, and the length of the shorter grating ($L$) is implemented as the distance between the mirrors.

All modeling was conducted using MATLAB, where $r_1$ and $\sqrt{A}$, were optimized numerically to achieve the best agreement with the experimental data. A comparison of the experimental and computational results is provided in **Figure S2b**, demonstrating single resonator performance for various $W$ and $L$ configurations.

Furthermore, the back-reflection of the resonator (*W=0.75, L=100*) was experimentally measured from the input port, as shown in **Figure S2c**. At the resonance wavelength, the total back-reflection was approximately -5 dB, primarily attributed to the resonator itself. However, the reflection spectrum indicates that at wavelengths outside the resonance, the dominant contribution to back-reflection originates from the grating couplers, measured at approximately -19 dB.

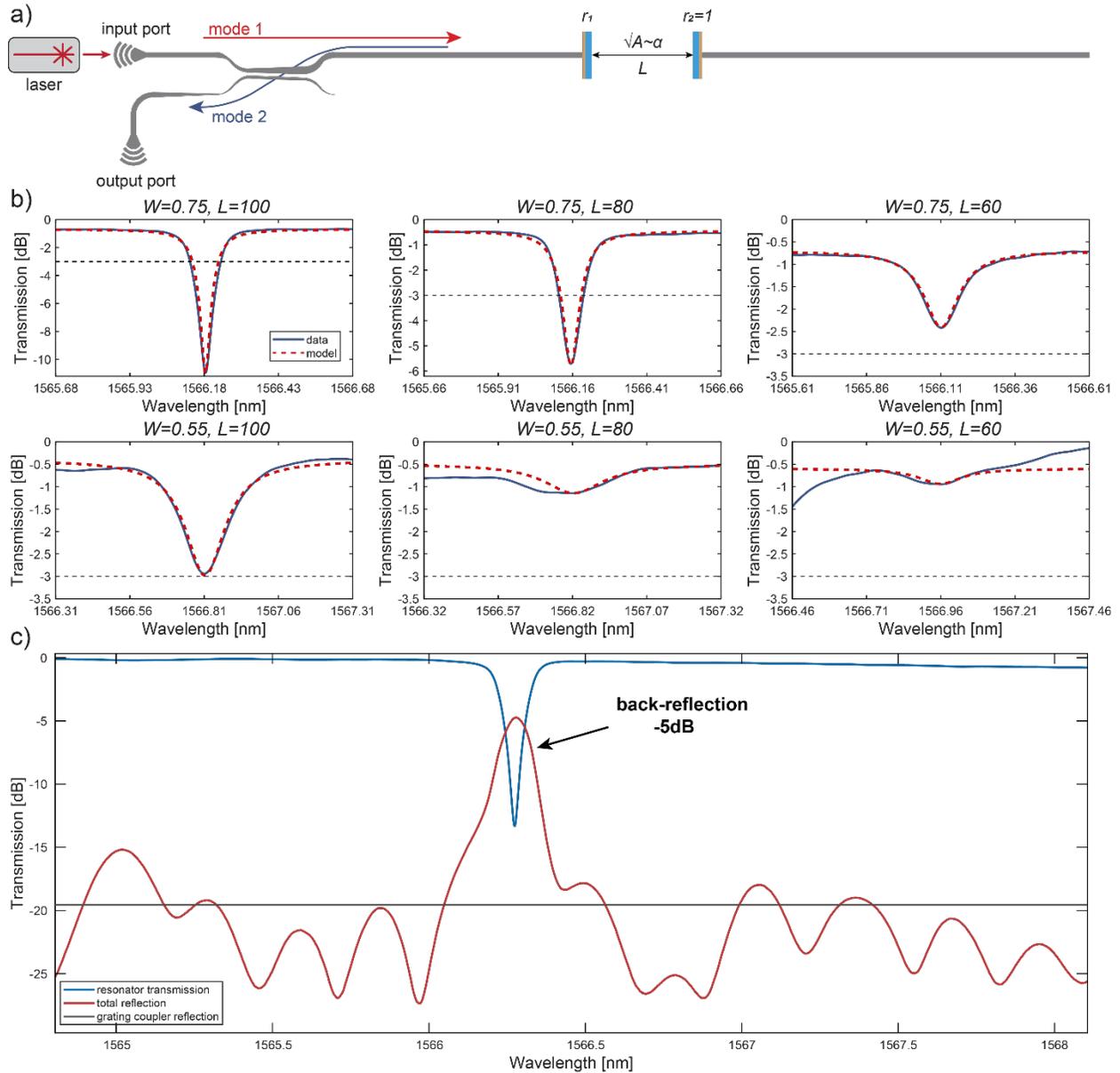

**Figure S2. Modelling of single resonators. a)** Schematic of the single resonator used for computational modelling. **b)** Computational and experimental transmission of the fabricated devices. **c)** Experimental transmission and back-reflection of the device (*W=0.75, L=100*).

The following equations are used for modelling:

$$r = -\frac{r_1 - Ae(-i\beta)}{1 - r_1 A \exp(-i\beta)} \quad (S1)$$

where

$$\beta = \frac{4\pi L n_{eff}}{\lambda} + \varphi_0 \quad (S2)$$

Equivalently,

$$E_{out} = -i \frac{r_1 - Ae^{\left(-i\left(\frac{4\pi L n_{eff}}{\lambda}+\phi_0\right)\right)}}{1 - r_1 Ae^{\left(-i\left(\frac{4\pi L n_{eff}}{\lambda}+\phi_0\right)\right)}} \quad (S3)$$

Here, $r, \beta, E$ represent the complex reflectivity of the resonator, the propagation constant within the cavity, and the complex electric field at the resonator's output, respectively.

Furthermore, to model the performance of these resonators under modulation, Equation *S2* is modified to account for the phase shift introduced during the modulation process.

$$\beta = \frac{4\pi L n_{eff}}{\lambda} + \varphi_0 + \frac{-2\pi L V}{d\,\lambda} \alpha n_{eff}^3 r_{33} \quad (S4)$$

Here, $V$ represents the applied voltage, d is the horizontal distance between the electrodes, $r_{33}$ denotes the electro-optic coefficient of the lithium niobate crystal, and $\alpha$ is the attenuation constant used to account for the decay of the generated external electric field during modulation as a function of the vertical distance between the electrodes and the waveguide.

To determine the coefficient $\alpha$, we designed an experiment in which only the amplitude of the applied signal was varied while recording the output of the MZI. By analyzing the relationship between the maximum output signal level and the input signal amplitude, we were able to choose an initial value for $\alpha$.

Equations *S1–S4* were employed to simulate the performance of the MZI devices under differential modulation. By comparing the simulated small-signal response during in-phase and out-of-phase modulation with the experimental data shown in **Figures 3a** and **3b** of the main text, we refined the value of $\alpha$ for greater accuracy. Using the simulation results, we calculated the phase shift achieved during differential modulation. Each resonator demonstrated a phase shift of $\frac{\pi}{90}$ when subjected to an applied voltage of $\pm 10$ V.

These simulations not only validated the proposed theoretical framework and successfully replicated the experimental results, but also provided deeper insights into the behavior of the modulators. For instance, they highlighted critical factors influencing device performance, such as the impact of resonance detuning, which is discussed in the following section.

## S3. Studying the Effects of Detuning on Small Signal Response

As discussed in the main text, resonance detuning significantly influences the small-signal response, impacting both its overall profile and key parameters such as isolation and probing wavelength. To investigate this effect further, simulations were conducted by varying the detuning to observe its effect on the small-signal response of the MZI. These simulations utilized the computational model developed for MZI$_1$, shown in **Figure 3a** of the main text. Detuning was controlled in the models by adjusting the initial phase ($\varphi_0$) of one of the resonators, as defined in **Equation *S4***.

**Figures S3a** and **S3b** illustrate the small-signal responses of MZI$_1$ for detuning values of $\Delta\lambda = 0.19\ nm$, $\Delta\lambda = 0\ nm$, respectively. A clear observation from these results is that the overall response profile undergoes drastic changes with detuning. For instance, the isolation values derived from the simulated responses were 38.7 dB and 30 dB for $\Delta\lambda = 0.19\ nm$, $\Delta\lambda = 0\ nm$, respectively. Additionally, the operating wavelength of the device shifted from 1566.49 nm to 1566.39 nm as the detuning decreased from 0.19 nm to 0 nm.

**Figure S3c** further quantifies the isolation as a function of resonance detuning. Although the average isolation value remains near 30 dB, certain detuning values result in significantly reduced isolation. In these instances, the amplitude of the output signal changes appreciably during both in-phase and out-of-phase modulations, thereby impairing the performance of the device as an IQ modulator. This undermines the intended functionality of achieving distinct amplitude and phase modulation.

Moreover, the operating wavelength of the modulators varies with detuning, necessitating adjustments to the probing wavelength to ensure optimal performance and maintain the desired isolation for practical applications. These findings emphasize the importance of precisely controlling resonance detuning in the design and operation of these modulators.

To ensure reliable operation and optimal performance of our devices, precise control of the detuning value is crucial for achieving the desired isolation. To address this, Ti/Au heaters were deposited on top of the longer Bragg gratings to actively tune the resonance wavelength of the resonators and compensate for detuning within the MZI device.

**Figure S4** highlights the performance of these integrated heaters. As illustrated, the heaters provide a resonance wavelength shift of up to 400 pm with approximately 0.2 mW of applied power. This tuning range is more than sufficient to compensate for the 0.2 nm detuning observed in fabricated devices, ensuring consistent and predictable device behavior.

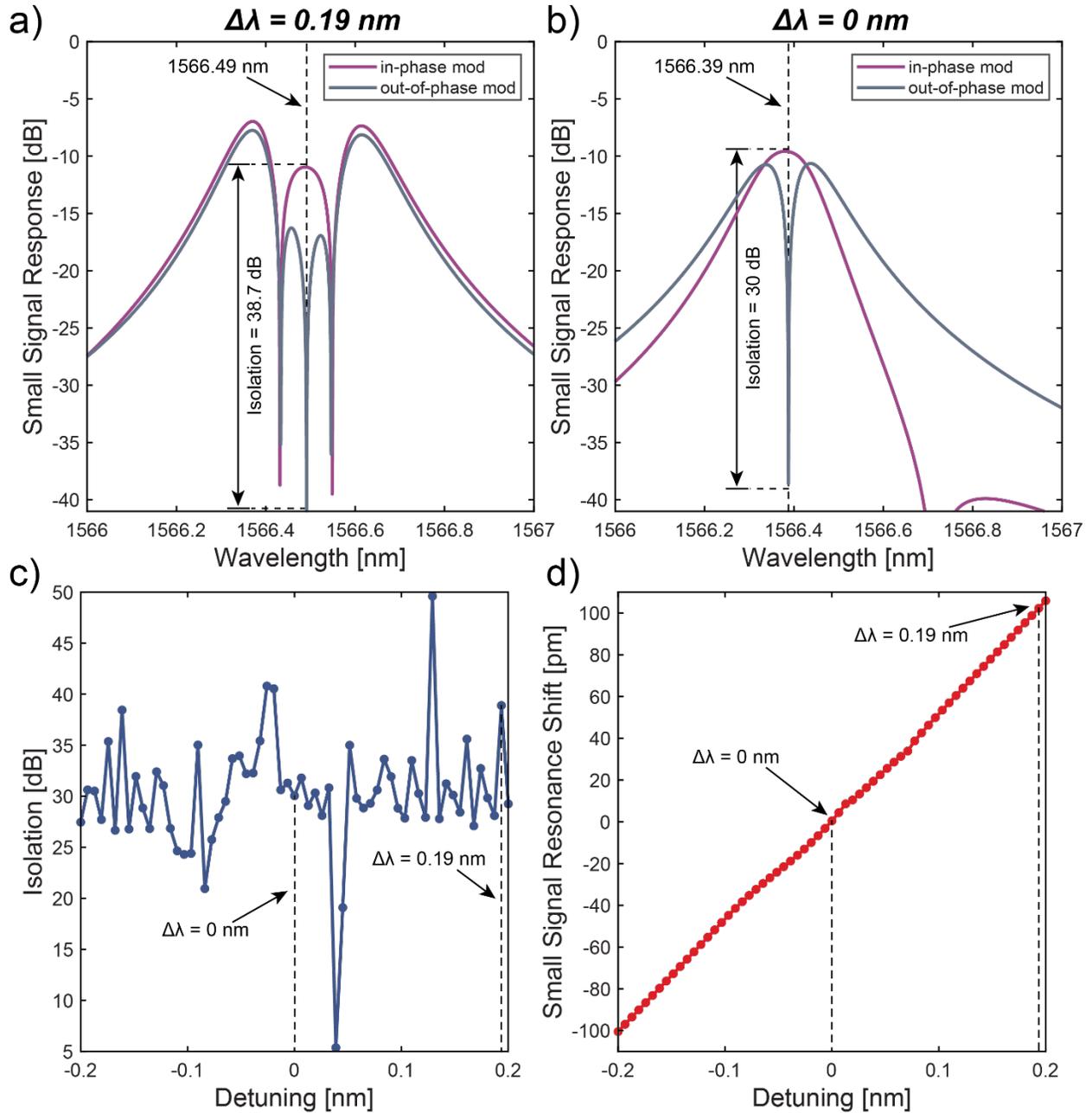

**Figure S3. Effect of detuning in small signal response of the coherent modulator. a-b)** Small signal response of MZI for resonance detuning values of $\Delta\lambda = 0.19\,nm$ **(a)**, and $\Delta\lambda = 0\,nm$ **(b)**. **c)** Changes in the isolation between in-phase and out-of-phase modulation as a function of detuning. **d)** Changes in the resonance wavelength of the channel as a function of detuning between single resonator pair.

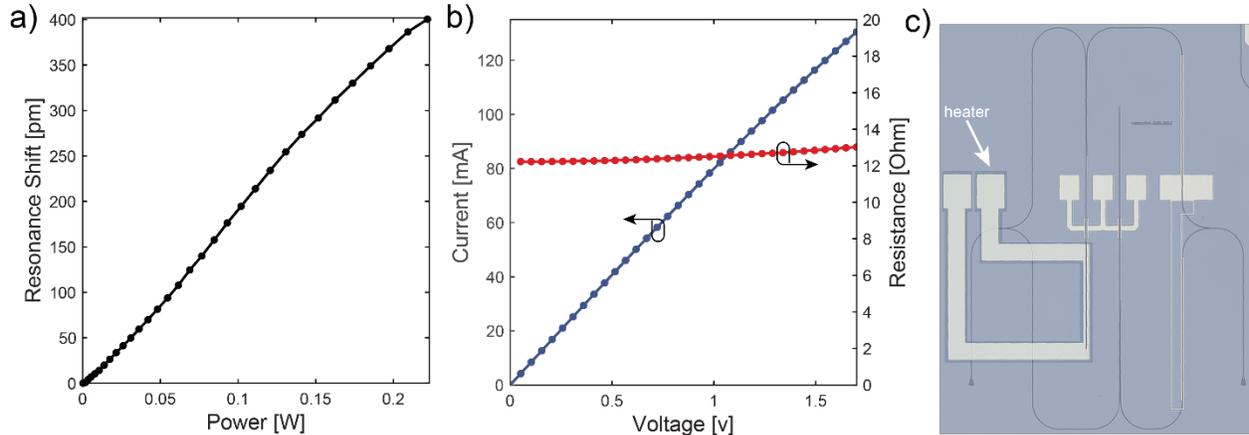

**Figure S4. Heater characterization. a)** Amount of resonant shift achieved as a function of heater power. **b)** I-V characteristics of deposited heaters. **c)** optical microscope image of the coherent modulator post heater deposition.

## S4. QAM Constellation

Quadrature amplitude modulation (QAM) is a widely adopted scheme in high-speed optical communication systems, combining amplitude and phase modulation to transmit higher amounts of information per symbol. In-phase (I) and quadrature (Q) modulators are essential components of QAM systems, enabling independent control of the optical signal's amplitude and phase. In this section, we evaluate the performance of our coherent resonant modulators for QAM applications.

While our modulators can achieve pure amplitude and phase modulations through in-phase and out-of-phase modulation schemes, simultaneous control of both amplitude and phase using symmetrical modulation schemes is not feasible. This limitation restricts the performance of these devices to 4-QAM at best. To overcome this challenge, we introduce an asymmetric modulation approach, which has the potential to enable more spectrally efficient modulations, such as 16-QAM. High-accuracy QAM constellations require precise and simultaneous control of both amplitude and phase.

In this context, by symmetrical modulation we refer to applying RF signals with identical amplitude levels (e.g., 10 volts) to the resonators. Conversely, asymmetric modulation involves applying RF signals with different amplitude levels to the resonators. This difference in amplitude levels induces asymmetric movement of the resonators during differential modulation. Such asymmetry provides a degree of control over the phase during amplitude modulation (in-phase modulation) and, similarly, some control over the output amplitude during phase modulation (out-of-phase modulation). By leveraging asymmetric modulation, our devices can potentially achieve higher-order QAM constellations, improving spectral efficiency in optical communication systems.

To better assess the performance of these MZIs, we considered three distinct scenarios in which only a portion of the complex modulation space was accessed during differential modulation. This limitation was introduced by constraining the modulation range of the resonators.

In the first scenario, we examined the case where the device achieved the full complex modulation space. For this, the modulation range of each resonator was restricted to phase shifts between $\pm\frac{\pi}{10}$. As shown in **Figure S5a**, this modulation range allowed the device to cover the entire complex plane. Computational results were obtained by performing a 2D sweep of the resonators' phases, and 2D interpolation was employed to generate the 16-QAM constellation. In the second scenario, we limited the modulation range of each resonator to $\pm\frac{\pi}{35}$, which resulted in the device covering only half of the complex plane. For the 16-QAM constellation, one half of the constellation was generated as described in the previous scenario. The second half was obtained by applying a $\pi$ phase shift to the MZI output (**Figure S5b**). In the third scenario, we restricted the modulation range of each resonator to $\pm\frac{\pi}{90}$, enabling the device to cover one quadrant of the complex space at a time. A 16-QAM constellation could then be generated by shifting the output phase of the modulator in integer multiples of $\pm\frac{\pi}{2}$ (**Figure S5c**). Depending on the modulation efficiency of the resonators, one of these approaches would be chosen to generate the QAM constellation.

As detailed in Supplementary **Section S2**, our resonators were able to achieve a modulation range of $\pm\frac{\pi}{90}$ with applied signals at ±10 Volts. Therefore, we selected the third approach to implement 16-QAM. Additionally, we explored the impact of varying the signal-to-noise ratio (SNR) on the quality of the generated constellation. This analysis aimed to understand how noise and signal distortion influenced the quality of the generated QAM constellation. **Figure S5d** illustrates the generated QAM constellations for SNR values of 30 dB, 20 dB, 15 dB, and 10 dB.

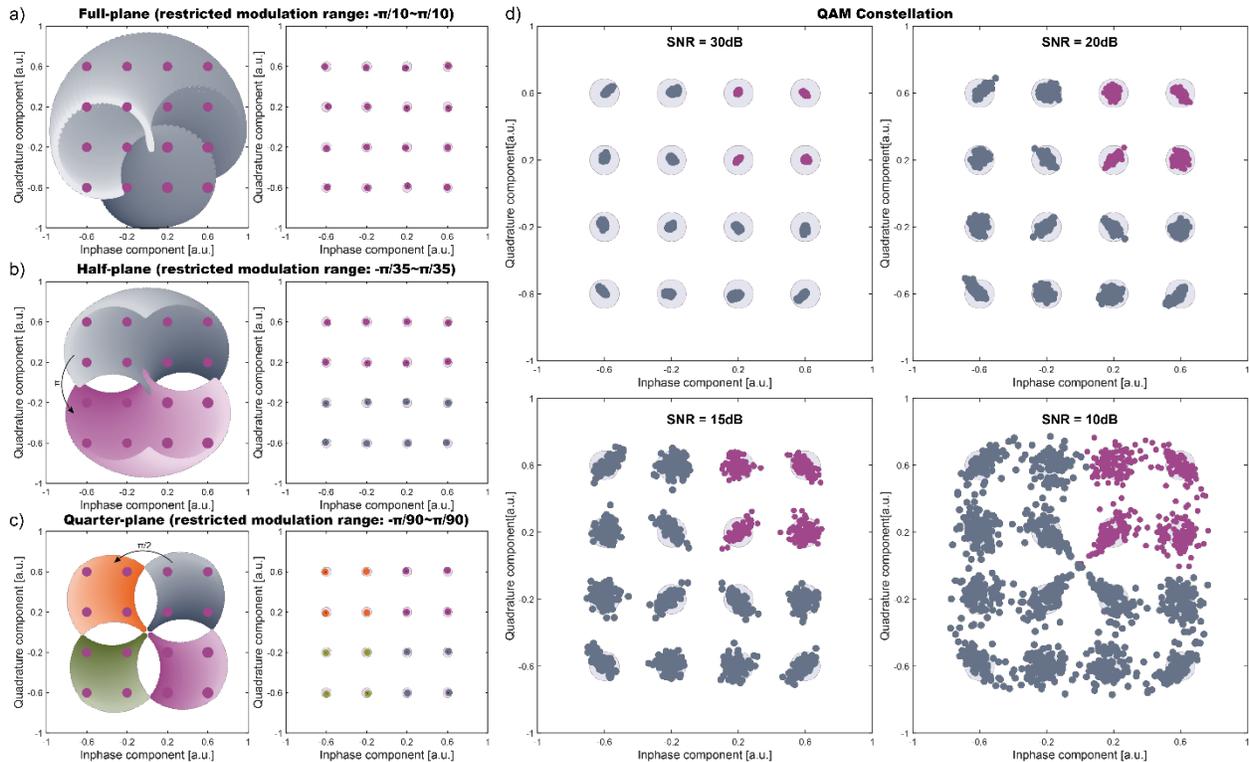

**Figure S5. Implementation of 16-QAM using coherent modulators. a-c)** achievable complex modulation space across various modulation ranges. **d)** Computationally generated QAM constellation using coherent modulator for different input SNR values.

Preliminary analysis indicates that implementing 16-QAM using our device is feasible and can be achieved through the methods outlined above. Our design demonstrates the capability to achieve a symbol rate of 29 Gbaud, which corresponds to an equivalent bit rate of 116 Gbit/s in digital systems. The relationship between the symbol rate and bit rate for 16-QAM is defined by the following equation:

$$bit\ rate = (symbol\ rate) \times \log_2 N \qquad (S5)$$

Where $N$ is the number of symbols, for 16-QAM, $N$ is equal to 16.